\documentclass{nature}

\title{Topologically trivial gap-filling in superconducting Fe(Se,Te) by one dimensional defects}

\author{A. Mesaros$^{1}$, G. D. Gu$^{2}$ \& F. Massee$^{1, *}$}

\usepackage{lineno}
\usepackage[english]{babel}
\usepackage{amssymb,amsmath}
\usepackage{graphicx}
\usepackage{multirow}
\usepackage{bigstrut}
\usepackage{xcolor}
\begin{document}
	
	\maketitle
	
	\begin{affiliations}
		\item Universit\'{e} Paris-Saclay, CNRS, Laboratoire de Physique des Solides, 91405, Orsay, France
		\item Condensed Matter Physics and Materials Science Department, Brookhaven National Laboratory, Upton, NY 11973, USA\\
		* freek.massee@universite-paris-saclay.fr
	\end{affiliations}
	
	\begin{abstract}
		Structural distortions and imperfections are a crucial aspect of materials science, on the macroscopic scale providing strength, but also enhancing corrosion and reducing electrical and thermal conductivity. At the nanometre scale, multi-atom imperfections, such as atomic chains and crystalline domain walls have conversely been proposed as a route to topological superconductivity, whose most prominent characteristic is the emergence of Majorana Fermions that can be used for error-free quantum computing. Here, we shed more light on the nature of purported domain walls in Fe(Se,Te) that may host 1D dispersing Majorana modes. We show that the displacement shift of the atomic lattice at these line-defects results from sub-surface impurities that warp the topmost layer(s). Using the electric field between the tip and sample, we manage to reposition the sub-surface impurities, directly visualizing the displacement shift and the underlying defect-free lattice. These results, combined with observations of a completely different type of 1D defect where superconductivity remains fully gapped, highlight the topologically trivial nature of 1D defects in Fe(Se,Te). 
	\end{abstract}
	
\section*{Introduction}
Structural defects have seen a resurgence of interest following the realization that their own topological nature allows them to host topologically protected in-gap states\cite{Jiang_Nature_Review_Physics_2023}, i.e. states that are robust to perturbations other than those that break the symmetry of the system. This insensitivity includes most types of random disorder, and as such makes topologically protected states a promising platform for quantum information technology \cite{kitaev_2003, nayak_rmp_2008}. The most obvious structural defect where topologically protected states may be found is the boundary between the material and the vacuum: the end-points of a 1D chain \cite{bernevig_chain_2013, vonoppen_chain_2013, yazdani_chain_2014, paaske_chain_2016, wiesendanger_chain_2018, pawlak_2019, Yazdani_2021}, the edge of a 2D island \cite{ojanen_2D_2015, Bode_Science_2016, cren_2D_2017, wiesendanger_2D_2019, liljeroth_2D_2020}, or the surface of a 3D bulk \cite{Zhang_2009}. Topologically protected states have been predicted and evidenced also on point-like, line-like and surface-like structural defects within the material or at their intersections with material boundaries, such as dislocations \cite{Ran2009}, disclinations \cite{Hughes2014}, grain boundaries \cite{Morpurgo2016}, stacking faults \cite{Queiroz2019}, and step edges \cite{Bode_Science_2016}. More recently, higher-order topology was discovered as a way to protect states at lower dimensional features of structural defects, such as corners on a surface \cite{bernevig_edge_2017, vonoppen_edge_2017, HOTIBi}, making a detailed understanding of the boundary essential. 
	
In the same way as a boundary with vacuum, the boundary between two materials may provide suitable conditions for Majorana modes to exist \cite{pawlak_2019, fu_kane_prl}. For a junction between two superconductors with a $\pi$-phase difference it was shown that dispersive Majorana modes can exist \cite{fu_kane_prl}. Recently, a relatively constant in-gap density of states on a structural defect line in crystalline Fe(Se,Te) was reported as evidence of such modes \cite{wang_science_2020}. This raises three important questions: 1) does a relative displacement occurring across a structural defect line induce a phase shift in the superconductor, 2) are the exact direction and magnitude of the structural displacement crucial for the appearance of the in-gap states, i.e. do small perturbations to the displacement lift the topological protection, and 3) what happens at the endpoint of a defect line.

In this work we measure single crystalline FeSe$_{0.45}$Te$_{0.55}$ with our 300~mK scanning tunnelling microscope\cite{massee_rsi_2018} and apply a tailored lattice-phase-shift analysis in order to address these questions and to uncover the exact origin of the structural displacement shift. We find that the lattice shift across the defect is typically near $\pi/2$ instead of quantised at $\pi$, making the quantised $\pi$-shift of the superconductor phase unlikely. Further, we find that the in-gap density of states becomes gapped on some segments of the defect, even though the lattice displacement shift remains constant on those segments, giving strong evidence against topological protection of the in-gap modes. Based on all our observations, plus the fact that we are able to reposition sections of a 1D defect with the STM tip, a natural explanation for the in-gap density of states and on average $\pi/2$ phase shift is the presence of sub-surface debris that warps the top Fe(Se,Te) layer(s) and induces non-topological gap filling.

\begin{figure}
	\centering
	\includegraphics[width=\textwidth]{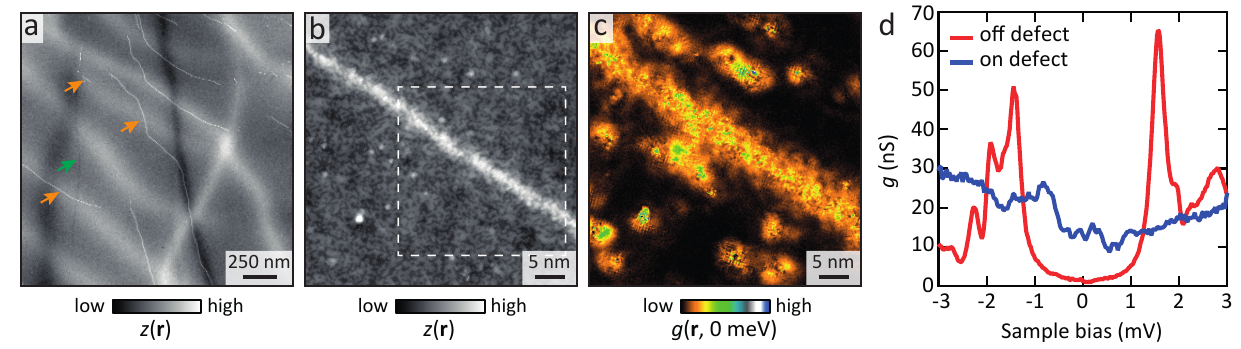}
	
	\caption{\label{fig:1} \textbf{1D defects in FeSe$_{0.45}$Te$_{0.55}$}. \textbf{a} Large field of view constant-current image of FeSe$_{0.45}$Te$_{0.55}$ containing a number of 1D defects (three are indicated by orange arrows). The green arrow marks a narrow dip that runs roughly vertically. Broader height variations likely reflect strain of the surface. $V$ = 50~mV, $I$ = 50~pA. The image has been corrected for several minor tip changes. \textbf{b} Constant current image on one of the 1D defects in \textbf{a}. $V$ = 5~mV, $I$ = 100~pA. The region in the dashed box is analysed in Fig. 2. \textbf{c} Differential conductance, $g$, at $E$ = 0 on the area of \textbf{b}. \textbf{d} Typical spectra taken on the 1D defect and several tens of nanometres away from it: on the defect the sub-gap states completely fill the gap.}
\end{figure}
	
\section*{Results}
\subsection{1D defects: topography and spectroscopy}
Large, atomically flat surfaces were obtained with clear atomic resolution. Additionally, a number of bright 1D defects were observed ranging in length from several tens to hundreds of nanometres. Interestingly, they do not have a fixed orientation with respect to the atomic lattice, are seen to change direction, and often abruptly terminate, see Fig.~\ref{fig:1}a. We note that in addition to these sharp 1D defects, broader intensity variations are also observed, which likely reflect a slight strain of the surface \cite{zhao_nphys_2021}. Lastly, a relatively narrow 1D dip defect can be seen (green arrow in Fig.~\ref{fig:1}a). 
	
\begin{figure}
	\centering
	\includegraphics[width=\textwidth]{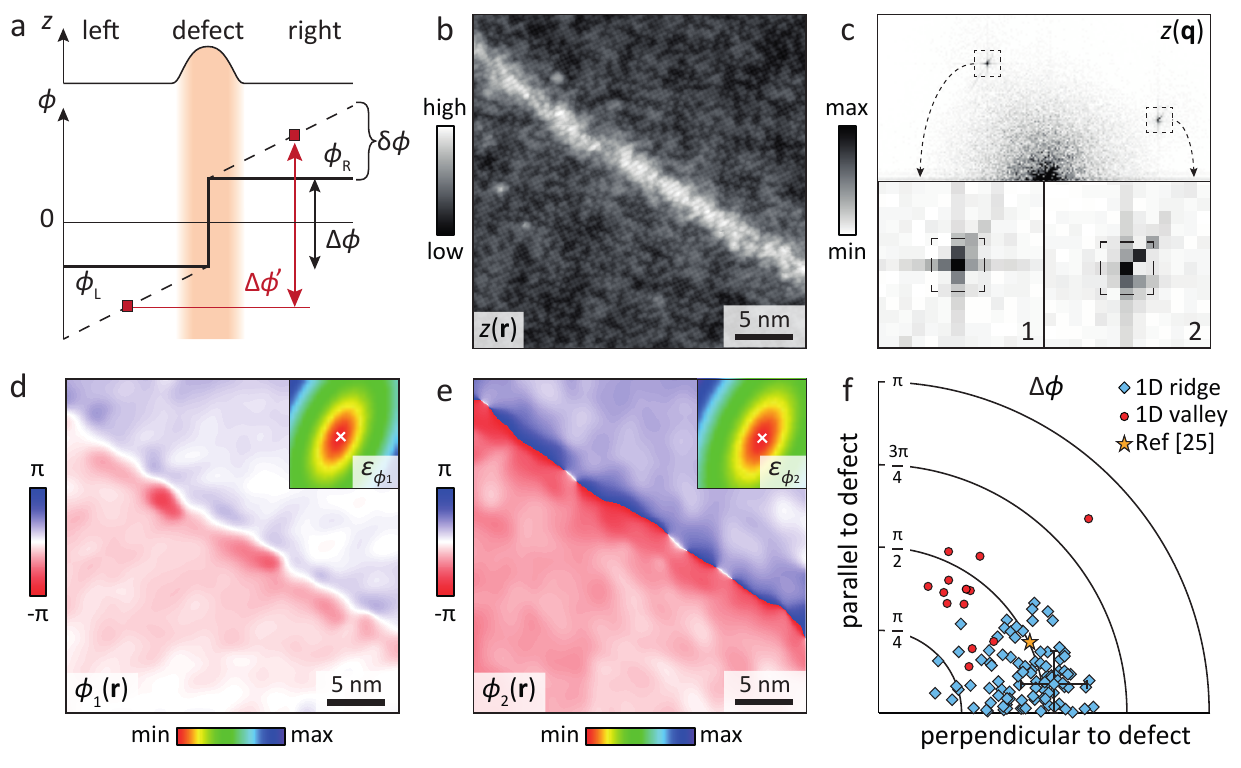}
	
	\caption{\label{fig:2} \textbf{Phase analysis}. \textbf{a} Schematic of the phase extraction method. Since atomic contrast in the defect region itself is blurred (orange shading), it cannot be used to extract the phase jump directly. Any non-zero slope in the phase (dashed line) due to a mismatch between the reference lattice constant and the actual lattice constant will introduce an error in the extracted phase: $\Delta\phi^{'} \neq \Delta\phi$. For an accurate determination of the phase jump, the slope (producing the phase accumulation $\delta \phi$) thus needs to be minimised. \textbf{b} Constant current image (dashed box of Fig.~\ref{fig:1}b). \textbf{c}  Fast Fourier transform of \textbf{b}, the insets show the two Bragg peaks used for the phase analysis. Note that Bragg peak 2 is clearly split. \textbf{d}, \textbf{e} Phase images from the two Bragg peaks. The insets show the total standard deviation of the phase ($\varepsilon_{\phi\alpha}$, see text) for reference lattices spanning the dashed areas of the insets of \textbf{c}. The phase images are the optimized ones obtained from the reference lattice which minimizes $\varepsilon_{\phi\alpha}$ (white cross). \textbf{f} Magnitude of the displacement shift vector $(\bar{\phi}^R_1-\bar{\phi}^L_1,\bar{\phi}^R_2-\bar{\phi}^L_2)$ between the two domains (see text) and its angle with respect to the 1D defect for a large number of 1D defects (both ridge- and dip-like). The phase shift of the ridge-like 1D defect (blue diamonds) is on average $\pi /2$ and predominantly oriented perpendicularly to the defect. The typical error bars, shown for the point extracted from panel b, predominantly reflect random phase fluctuations on both sides of the defect (see also Supplementary Note 2).}
\end{figure}
	
Upon closer inspection, the bright 1D defects (orange arrows in Fig.~\ref{fig:1}a) show identical characteristics to those reported previously \cite{wang_science_2020}: the defects are a continuous protruding line in topography a few nanometres wide, lacking clear atomic contrast, see Fig.~\ref{fig:1}b. Additionally, the in-gap density of states along the entire length of the defect is enhanced, effectively filling up the gap, see Fig.~\ref{fig:1}c,d. We note that despite the presence of a small concentration of excess Fe atoms (whose distribution is the same near and away from 1D defects), the tunnelling spectra away from the defect have a well defined gap with minimal sub-gap filling and gap sizes in agreement to those reported previously \cite{hanaguri_science, massee_sciadv, wang_science_2020}. 
	
\subsection{Lattice phase shift determination}
Next we discuss the displacement of the lattice upon crossing the 1D defect. With respect to the ideal square lattice, the displacement vector $(u_x,u_y)=\frac{a}{2\pi}(\phi_1,\phi_2)$ of each atom is defined within the periodic unit-cell of sidelength $a$, and is hence represented by two phase variables $\phi_\alpha\in[0,2\pi]$, $\alpha=1,2$, called the lattice phases. As detailed in Refs. [\citenum{hytch_ultramicroscopy_1998, lawler_fujita}] the smoothed lattice phases $\phi_\alpha(\mathbf{r})$ defined at all positions $\mathbf{r}$ of the topographic image are determined by comparing the atomic contrast in the topograph with a reference lattice. To accurately extract the lattice phase, the choice of reference lattice is critical. This is because if the position $\mathbf{K}_\alpha$ of the Bragg peak of the reference lattice slightly deviates from the one of the measured lattice, an extracted lattice phase image $\phi_\alpha(\mathbf{r})$ will have a non-zero linear term, i.e. a slope, since $\cos[\sum_\alpha(\mathbf{K}_\alpha+\delta \mathbf{K}_\alpha)\cdot \mathbf{r} +\phi_\alpha(\mathbf{r})]=\cos[\sum_\alpha \mathbf{K}_\alpha\cdot\mathbf{r}+\phi_\alpha(\mathbf{r})+\delta\phi_\alpha(\mathbf{r})]$, with the linear term $\delta\phi_\alpha(\mathbf{r})\equiv\delta \mathbf{K}_\alpha\cdot\mathbf{r}$. If the slope term is nonzero for a given choice of reference lattice, one will over- or underestimate the relative displacement of domains (labeled left ($L$) and right ($R$)), that may be present across the 1D defect, see Fig.~\ref{fig:2}a. An operational definition of the ideal Bragg peak $\mathbf{K}_\alpha$ is therefore that for which there is a vanishing slope of the $\phi_\alpha(\mathbf{r})$ image \cite{Mesaros2016}. Crucially, since the pixel size of the experimental image is generally not commensurate with $a$, the $\mathbf{K}_\alpha$ is not necessarily positioned on the center of a pixel in the Fourier transform of the topograph \cite{hytch_ultramicroscopy_1998, lawler_fujita, Mesaros2016}. 
	
\subsection{Reference lattice optimisation}
To determine the optimal reference lattice, we calculate the lattice phase images $\phi^{\mathbf{K}_\alpha}_{\alpha}(\mathbf{r})$ for a finely spaced set of values of $\mathbf{K}_\alpha$ covering the experimental 4x4 pixel area centred on the brightest pixel in the Fourier transform (Fig.~\ref{fig:2}c). In practice, each phase image is obtained by shifting the Fourier data so that $\mathbf{K}_\alpha$ is at the origin, then applying a low-pass filter and inverse Fourier transforming. As such, the image visualizes the local displacements of the measured topography with respect to the reference lattice defined by $\mathbf{K}_\alpha$, exactly as detailed in e.g. Refs.~[\citenum{lawler_fujita, wang_science_2020}]. The crucial difference with previous work, however, is that we allow for non-integer values of $\mathbf{K}_\alpha$. Then, for each $\phi^{\mathbf{K}_\alpha}_{\alpha}(\mathbf{r})$ image (Figs.~\ref{fig:2}d,e), we calculate the total standard deviation $\varepsilon_{\phi\alpha}(\mathbf{K}_\alpha)\equiv(N_L\varepsilon^L_{\phi\alpha}(\mathbf{K}_\alpha)+N_R\varepsilon^R_{\phi\alpha}(\mathbf{K}_\alpha))/(N_L+N_R)$ from the standard deviations $\varepsilon^{L/R}_{\phi\alpha}(\mathbf{K}_\alpha)$ and number of pixels $N_{L/R}$ of the real-space domains $L$ and $R$, respectively (insets of Figs.~\ref{fig:2}d,e). The minimum value of the total standard deviation $\varepsilon_{\phi\alpha}(\mathbf{K}_\alpha)$ over the set of considered $\mathbf{K}_\alpha$ then determines the optimal $\mathbf{K}_\alpha$ of the reference lattice and its lattice phases (Figs.~\ref{fig:2}d,e), since a non-vanishing slope term always increases the standard deviation of a function on a domain. The key method change with respect to Ref.~[\citenum{Mesaros2016}] is that although a considered $\mathbf{K}_\alpha$ is fixed for the entire image, the standard deviations of the two domains $L,R$ are separately determined and then added up - without considering the region where the phase jump occurs. This is because we are aiming to determine the periodicity of the ideal lattice, which we assume to be identical on either side of the defect (see also Supplementary Note 2). Including the phase jump $\Delta\phi$ (which we cannot locally extract due to absence of atomic contrast) would erroneously increase the total slope of the phase across the image by $\Delta\phi/l$, where $l$ is the image length, thereby effectively stretching or compressing the resulting reference lattice with respect to the ideal one. We find that the lattice phases obtained with the ideal reference lattice are uniform on each of the domains $L,R$ in absence of drift during the experiment (see Figs.~\ref{fig:2}d,e and Supplementary Information section 1 and 2). Hence the displacement phase shift $(\Delta\phi_{1},\Delta\phi_{2})$ across the defect is accurately given by the difference of the phase averaged on each domain, $\Delta\phi_{\alpha}\equiv\bar{\phi}^R_\alpha-\bar{\phi}^L_\alpha$. Additionally, we can extract the angle of the 1D defect with respect to the atomic lattice, and thus the angle between the displacement shift and the 1D defect. Fig.~\ref{fig:2}f shows the magnitude of the displacement phase, $\Delta\phi\equiv\sqrt{(\Delta\phi_{1})^2+(\Delta\phi_{2})^2}$, and its angle to the defect extracted for nearly 100 topographies taken on different locations of different (ridge-like) 1D defects in Fig.~\ref{fig:1}a. Two observations stand out: the phase shift magnitude is on average $\pi/2$ (corresponding to a quarter of a unit-cell), and the displacement tends to be perpendicular to the 1D defect. We note that the data of Ref.~[\citenum{wang_science_2020}] falls perfectly on top of our data when the above phase-slope optimization method is applied (we recover their sloped results using integer pixel reference lattices). For comparison, we performed the same analysis on several locations on a valley-like 1D defect (green arrow of Fig.~\ref{fig:1}a), which has the same phase shift magnitude, but has its orientation along the defect instead of perpendicular to it, see also Supplementary Information section 3. Finally, we note that the same displacement phase shifts are obtained if the real-space fitting method introduced by Ref.~[\citenum{pasztor_prr_2019}] is applied, see Supplementary Information section 1.
	
\begin{figure}
	\centering
	\includegraphics[width=\textwidth]{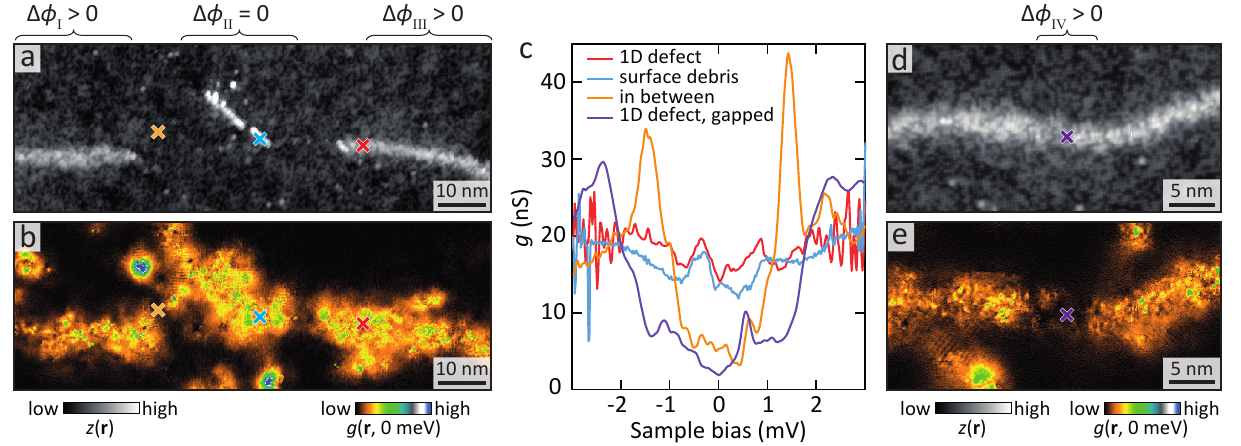}
		
	\caption{\label{fig:3} \textbf{End points, debris and interruptions}. \textbf{a} Constant current image of a split 1D defect. In between the two segments is a strip of (1D) surface debris. $V$ = 5~mV, $I$ = 100~pA. \textbf{b} $g$($E$ = 0) taken simultaneously with \textbf{a}. Even though there is within error no phase shift of the lattice at the surface debris area ($\Delta \phi_{\mathrm{II}}$ = 0.14$\pi\ \pm$ 0.19$\pi$), the in-gap density of states is indistinguishable from that of the 1D defect (that has non-zero phase shifts $\Delta \phi_{\mathrm{I}}$ = 0.52$\pi\ \pm$ 0.12$\pi$ and $\Delta \phi_{\mathrm{III}}$ = 0.54$\pi\ \pm$ 0.16$\pi$). \textbf{c} Spectra marked with crosses in \textbf{a} on the 1D defect, the surface debris and in between the two. \textbf{d} Constant current image and \textbf{e} $g$($E$ = 0) on another 1D defect. Setup: $V$ = 5~mV, $I$ = 80~pA. Even though the phase shift of the lattice is constant along the 1D defect ($\Delta \phi_{\mathrm{VI}}$ = 0.38$\pi\ \pm$ 0.25$\pi$), the differential conductance shows a gap. The cross marks the location of the spectrum in \textbf{c}.}
\end{figure}
	
\subsection{Surface debris, endpoints and gapping}
The seeming absence of a $\pi$ lattice phase shift poses a challenge: could a non-quantized and hence non-topological value of the lattice phase shift still induce a quantized $\pi$ phase shift in the superconductor, and, if not, is there an alternative explanation for the near-constant in-gap density of states along the 1D defect? While searching for an answer to these questions, we encountered several examples of unexpected behaviour. The first one is the presence of debris that forms 1D-like structures on top of the surface, see Fig.~\ref{fig:3}a. Although the surface debris is not accompanied by a phase shift of the lattice, it shows nearly identical behaviour in differential conductance (Fig.~\ref{fig:3}b, c and Supplementary Information section 4). Interestingly, the surface debris is often located close to an endpoint of a 1D defect, suggesting a possible link. Lastly, on continuous 1D defects, with a phase shift of order $\pi/2$ along the length of the defect, we find several short segments with a recovered gap in the density of states, see Fig.~\ref{fig:3}d, e and Supplementary Information section 5. Similarly, the dip-like 1D feature in Fig.~\ref{fig:1}a, which has a phase shift of similar magnitude as the ridge-like 1D defect, has little to no in-gap filling, see Supplementary Information section 3.
	
\subsection{Manipulating 1D defects}
Before discussing the implications of these findings, we first focus on the surface debris and its possible relation to the 1D defects. One of the challenges with any type of adatom or debris is that it is relatively easily moved and/or picked up by the STM tip. During our studies, in particular of the area in Fig.~\ref{fig:3}a, it proved very difficult to complete a spectroscopic map without modifying the debris and/or tip, suggesting that the debris is rather sensitive to the electric field of the tip. Curiously, we also observed small instabilities in the tunnelling signal on parts of the ridge-like 1D defects, although the tip and surface themselves did not seem to be modified, see Supplementary Information section 6 for examples. 
	
\begin{figure}
	\centering
	\includegraphics[width=\textwidth]{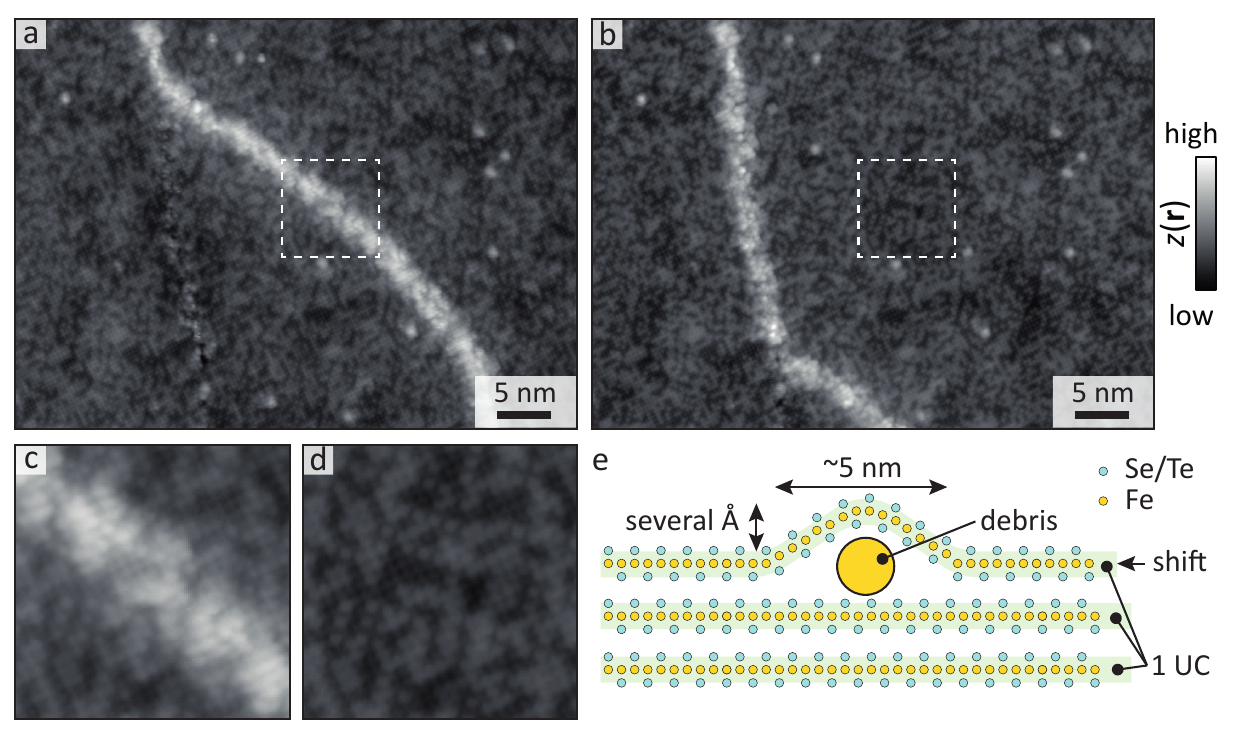}
		
	\caption{\label{fig:4} \textbf{Defect manipulation}. Constant current image of a 1D defect before (\textbf{a}) and after (\textbf{b}) manipulating it with the electric field of the tip. $V$ = 5~mV, $I$ = 80~pA. \textbf{c}, \textbf{d} Area in the dashed boxes of \textbf{a}, \textbf{b}, i.e. with and without defect, respectively. All atoms are clearly accounted for, they are simply pushed upwards in \textbf{c}. \textbf{e} Schematic of the 1D defect: debris (likely Fe clusters) underneath the surface warps the top unit cell (UC) layer, leading to a small shift of the lattice. The size of the debris is exaggerated for clarity.}
\end{figure}
	
To explore the instabilities in more detail, we scanned at increasingly smaller junction resistances (i.e. smaller tip-sample distance meaning higher electric field) on a 1D defect where instabilities occurred. Occasionally, the 1D defect seemed to slightly change orientation before snapping back, until at sufficiently low junction resistance it was permanently displaced to a new location, see Fig.~\ref{fig:4}a, b. Subsequent treatments enabled us to relocate additional sections of the 1D defect. The differential conductance before and after manipulation clearly shows the strong link between the defect and the enhanced in-gap signal (see Supplementary Information section 6 for more details). More importantly, we can now resolve the atomic lattice where the 1D defect used to be, as well as directly confirm that the lattice shifts only across the 1D defect. Interestingly, as Figs.~\ref{fig:4}c, d show, the lattice on top of the 1D defect and the same area after moving it away look very similar. In fact, all surface atoms seem to be accounted for, they are simply somewhat closer together on the 1D defect. Combined with the elevated height of the 1D defect, this strongly suggests that the surface is warped over a sub-surface defect as schematically depicted in Fig.~\ref{fig:4}e. This naturally explains the phase shift of the lattice and its preferential direction: it shifts perpendicularly towards the defect to accommodate the warping. Using the observed height and width of the defect, the lateral shift is estimated to be on the order of an \r{A}ngstrom, which is the same order of magnitude as the shift and corresponding phase difference observed in experiment.
	
All findings combined strongly suggest that the 1D defects are debris buried just below the surface (Fig.~\ref{fig:4}e). Although the nature of the debris is hard to determine, a possible candidate is Fe aggregations. Fe atoms are known to create long chains on Pb \cite{yazdani_chain_2014} under conditions that may also have been met during the growth of our single crystals.
	
\section*{Discussion}
Can surface and/or sub-surface debris host 1D Majorana modes? From our analysis, one of the essential requirements for 1D dispersing Majorana modes, namely a $\pi$-phase shift of the superconductor, is unlikely met: the lattice itself does not have such a quantized topological shift, so there is no reason to believe the superconductor would. Moreover, the gaps in density of states on small portions of a defect with a near constant lattice phase shift (Fig.~\ref{fig:3}d, e) strongly argue against the presence of topologically protected modes. If the debris has the correct spin texture, however, it may host Majorana bound states at the ends of the defects \cite{bernevig_chain_2013, vonoppen_chain_2013, paaske_chain_2016}, whereas the defect itself would host conventional Yu-Shiba-Rusinov bands \cite{schneider_naturenano_2022}. We find, however, no evidence of unusual behaviour at the endpoints of our 1D defects, see Fig.~\ref{fig:3}a and Supplementary Information section 7, suggesting that Majorana bound states are unlikely to be present. We therefore conclude that the gap filling at these 1D defects is topologically trivial in origin: strong scattering at clusters of impurities suppresses superconductivity and fills up the sub-gap region with states. The observation that local damage from heavy ion irradiation shows remarkably similar gap filling to that observed on 1D defects \cite{massee_sciadv} further indicates that this is not an unreasonable scenario. More generally, our results show that a roughly constant sub-gap density of states is not a unique signature of linearly dispersing modes of a topological superconductor, but in fact quite common for conventional mechanisms of gap-filling. Future studies at genuine crystalline domain walls may shed more light on whether or not topologically protected sub-gap states can truly be hosted by Fe(Se,Te).	
	
	\begin{methods}
		Fe(Se,Te) single crystals were grown using the self-flux method. As-grown samples with a superconducting transition temperature of 14.5~K were used throughout this work. The crystals were mechanically cleaved in cryogenic vacuum at T $\sim$ 20~K and directly inserted into the STM head at 4.2~K. An etched tungsten tip was used for all measurements. Differential conductance measurements were performed by numerical derivation as well as with a lock-in amplifier operating at 429.7~Hz. All measurements were recorded at the base temperature of T = 0.3~K.		
	\end{methods}
	
	\begin{addendum}
		\item[Data Availability] All raw data generated during the study are available from the corresponding authors upon request.
	\end{addendum}

	\begin{addendum}
		\item[Acknowledgements] FM would like to acknowledge funding from the ANR (ANR-21-CE30-0017-01). The work at BNL was supported by the US Department of Energy, office of Basic Energy Sciences, contract no. DOE-sc0012704.
		\item[Author Contributions] F.M. performed the experiments. F.M. and A.M. analysed the data, interpreted the results and wrote the manuscript. G.D.G. provided the samples.
		\item[Competing Interests] The authors declare that they have no competing interests.
	\end{addendum}
	
	\newpage
	
	\Large\bfseries\noindent\sloppy \textsf{Supplementary Information to `Topologically trivial gap-filling in superconducting Fe(Se,Te) by one dimensional defects'}
	
	\normalsize\normalfont
	
	\noindent\sloppy A. Mesaros$^{1}$, G. D. Gu$^{2}$ \& F. Massee$^{1}$
	
	\begin{affiliations}
		\item Universit\'{e} Paris-Saclay, CNRS, Laboratoire de Physique des Solides, 91405, Orsay, France
		\item Condensed Matter Physics and Materials Science Department, Brookhaven National Laboratory, Upton, NY 11973, USA
	\end{affiliations}
	
	\setcounter{page}{1}
	\setcounter{figure}{0}

\renewenvironment{figure}{\let\caption\figcaption}{}

\newcommand{\figcaption}[2][]{%
	\refstepcounter{figure}
	\ifthenelse{\value{figure}=1}{
	}{
}
\sffamily\noindent\textbf{Figure S\arabic{figure}}\hspace{1em}#2}

	\section{Integer $\mathbf{K}_\alpha$, arbitrary $\mathbf{K}_\alpha$ and real-space fitting analysis}	
\begin{figure}
		\centering
		\includegraphics[width=0.8\textwidth]{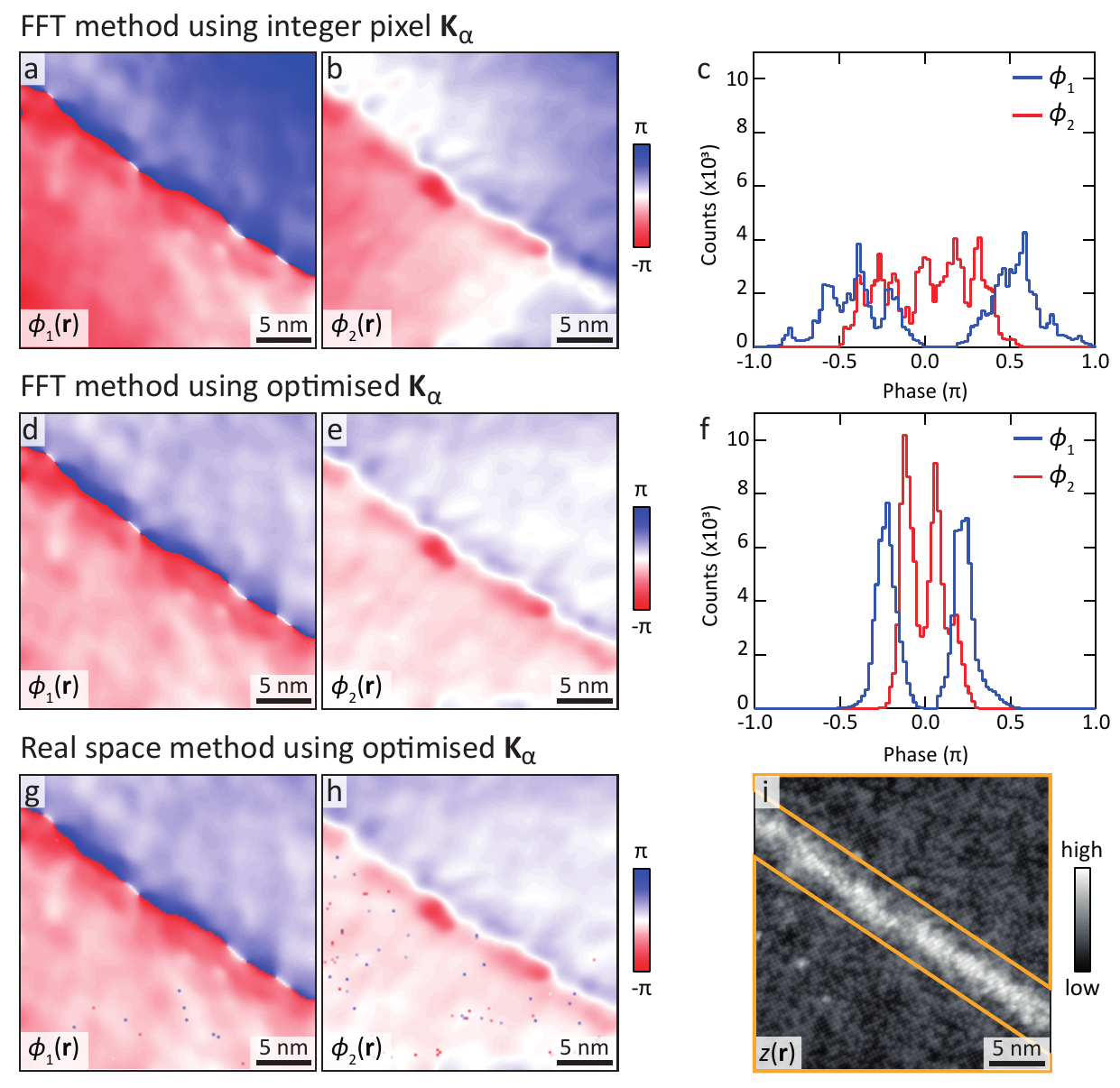}
		
		\caption{\textbf{Comparison of techniques.} \textbf{a, b} Phase images for the same 1D defect as main text Fig.~2b but using an integer pixel value for the $\mathbf{K}_\alpha$: ($K_{1x}$, $K_{1y}$) = (245, 150), ($K_{2x}$, $K_{2y}$) = (153, 120). \textbf{c} Histograms of \textbf{a}, \textbf{b} considering only the regions where the atomic contrast is not obscured by the 1D defect (see below). \textbf{d-f} Same as \textbf{a-c}, but using the optimal $\mathbf{K}_\alpha$: ($K_{1x}$, $K_{1y}$) = (244.65, 150.27), ($K_{2x}$, $K_{2y}$) = (152.67, 120.0). In this case, there is no slope in the phase images and the phase shifts across the 1D defect can be accurately extracted from the histograms. \textbf{g, h} Phase images determined using the real-space fitting method with the optimal $\mathbf{K}_\alpha$. The results are identical to \textbf{d}, \textbf{e} except for a few pixels where the real-space fit did not converge properly. \textbf{i} Constant current image that was used to extract the phase images in this figure. The orange lines mark the regions considered for the histogram and phase slope minimization, i.e. the area between the lines where the 1D defect is located is not considered as the atomic contrast is obscured here.}\label{fig:s_realspace}
	\end{figure}

Throughout this work the phase shifts in the atomic lattice are extracted using Fourier transform analysis. We refer to Ref. [\citenum{lawler_fujita}] for a detailed description of the analysis technique. We stress that we use the exact same technique as this work and e.g. Ref. [\citenum{wang_science_2020}], but extend the analysis to include non-integer pixel values of the reference lattice Bragg vector. This is essential because the experimental pixel size is generally not commensurate with the atomic lattice, hence a Bragg peak is in practice never at integer pixel coordinates. The optimization method we introduce based on minimizing the slope term in the phase field recovers an optimal (fractional) coordinate of the reference lattice vector, $\mathbf{K}_\alpha$. Fig.~S\ref{fig:s_realspace} demonstrates the difference in the extracted phase fields when the optimal $\mathbf{K}_\alpha$ and the closest integer-valued $\mathbf{K}_\alpha$ are compared. The histogram of phase values clearly shows that the optimal (fractionally-valued) $\mathbf{K}_\alpha$ reveals quite homogeneous domains and the rigid shift between them, while the closest integer-valued $\mathbf{K}_\alpha$ introduce large unphysical spatial fluctuations. We note that the histograms only consider the parts of the image where the atomic contrast is not obscured by the 1D defect (see Fig.~S\ref{fig:s_realspace}i). The distance between the respective peaks therefore gives an accurate value for the phase shift across the 1D defect and is used throughout this work.
	
	Another technique introduced by Ref. [\citenum{pasztor_prr_2019}] obtains phase information through a real-space fitting routine. To ensure that our Fourier-transform approach does not introduce artefacts or systematic errors, we have applied this technique as well to several of our topographies. Since the real-space method calculates the phase fields but does not determine  $\mathbf{K}_\alpha$, we input the optimal values of $\mathbf{K}_\alpha$ which we obtained from our Fourier transform analysis. After applying the same Gaussian filtering window as in the Fourier transform analysis, and then inverse Fourier transforming, we obtain the two real-space images (one for each Bragg peak) that will be used for fitting. For the real-space fit we select a sliding window of roughly one wavelength, and use a roughly 50\% overlap between subsequent windows. We note that similarly to Ref. [\citenum{pasztor_prr_2019}] the exact size of the window does not significantly influence the outcome of the fit as long as it is not much smaller or much larger than one wavelength. As the comparison in Figs.~S\ref{fig:s_realspace}g,h shows, both methods give identical results, which is strong support for the validity of our analysis. The main drawback of the real-space fitting method is that it is very time-consuming since a large number of real-space fits are required, of which some do not easily converge due to e.g. limited local contrast.
	
	\section{Phase shift in presence of non-linear strain or drift}
	
\begin{figure}
		\centering
		\includegraphics[width=\textwidth]{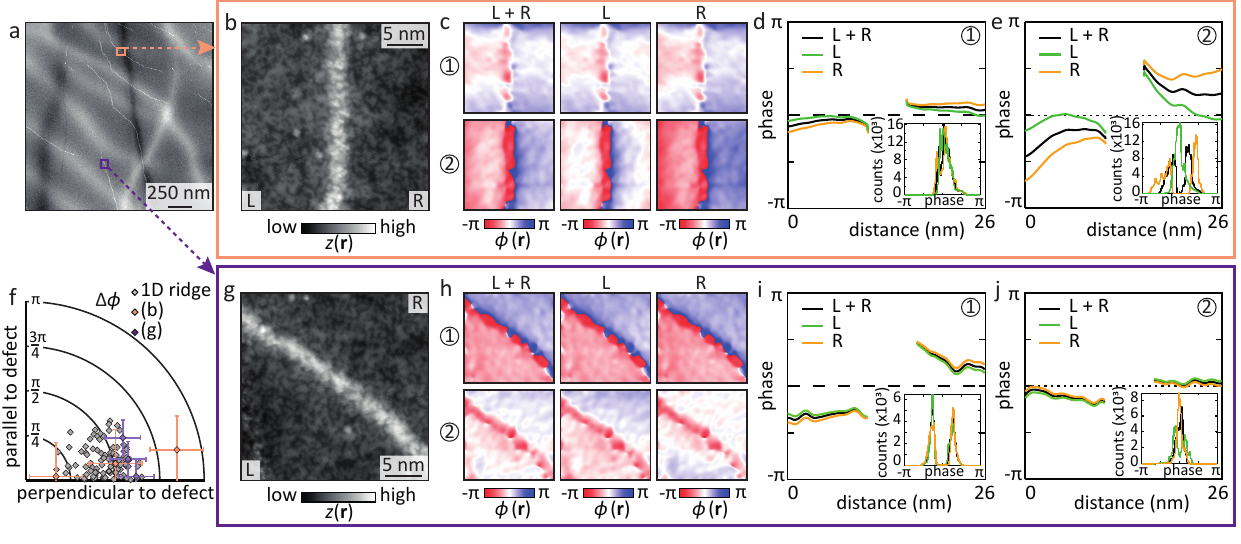}
		\caption{\textbf{Phase analysis with(out) strain.} \textbf{a} Large field of view (same as main text Fig.~1a). \textbf{b} Constant current image of 1D defect nearby the large scale dark depression in panel a, which is likely due to strain. \textbf{c} Phase analysis for the two  lattice vector directions (1 and 2) upon optimising the reference lattice using both sides of the defect ($L$+$R$), or either the left ($L$) or right ($R$) side only. \textbf{d, e} Horizontal line-cut through the middle of the phase images in panel c for lattice vectors 1 and 2, respectively. \textbf{f} Phase jump and its direction with respect to the defect for the three cases ($L$+$R$, $L$ and $R$, orange markers). To avoid under- or overestimating the phase jump in presence of strain, both sides need to be taken into account. \textbf{g-j} Same as b-e on a region without strain. The resulting phase jump magnitude for the three cases is near identical (purple markers in f).}\label{fig:s_strain}
	\end{figure}

	In this work we are purely interested in the phase jump across the 1D defect and would like to avoid or compensate for possible additional position-dependent changes in the phase. In principle, the lattice in absence of defects has a single lattice constant. We have confirmed this on a number of topographs taken far from 1D defects. Since there is no reason for either side of the 1D defect to be different, we can safely assume that the 1D defect only locally distorts the phase. To determine the optimal lattice parameter, we therefore combine the standard deviation of both sides of the 1D defect. In most cases, using only one of the sides of the 1D defect to determine the reference lattice will give near-identical results (see Fig.~S\ref{fig:s_strain}g-j).
	
	One may wonder, however, if there are changes in the phase possible that could affect the reliability of our phase jump extraction method, or where perhaps only one side of the 1D defect should be considered. Since we optimize the reference lattice parameter, only non-linear changes to the lattice constant are important: a linear compression or expansion will merely produce a slightly different reference lattice parameter, but still a constant phase across an image (and thus reliable jump). There are two possibilities for non-linearity: drift and strain. Drift, due to the slow (temperature dependent) relaxation of the piezos after a change in voltage, will stretch or compress an image in the direction of the drift. In this work we have tried to minimize non-linear drift by staying in the same location for extended periods of time, waiting sufficiently long after a temperature change and scanning slowly. Nevertheless, non-linear drift could still appear, for example in the first image after a temperature change. This, however, always shows up at the start of the image and in the slow scan direction (the y-axis in all our data) and as such can easily be recognized. For vertically running 1D defects, using both sides of the 1D defect is therefore valid this case. Nevertheless, whenever we did observe non-linear drift, we have not taken the affected part of the image into consideration. 
	
	A more interesting case is when the lattice constant itself is changing non-linearly as function of position due to strain. Again we stress that a linear strain field will be corrected for by our optimization method. A non-linear strain field, however, will give a different slope of the phase for the left and right region of the image. Let us consider a situation where the lattice parameter is constant on the left, has a positive phase jump to the right and then progressively compresses on the right (other configurations will lead to the same conclusion). Optimizing the reference lattice using the left side only will give a constant phase on the left, a positive jump followed by a phase with a negative slope on the right. Clearly, this will lead to an underestimation of the jump in the phase, as the negative slope partially counteracts the positive jump. Conversely, if the lattice is optimized on the right side only, the jump will be overestimated. Figure S\ref{fig:s_strain}b-e shows an example of exactly such a situation. Here, the 1D defect is close to a large scale depression in topography, likely associated with strain in the sample. As the analysis using either the left side, right side or both shows, the most reliable phase jump is extracted when taking both sides into account, as we do throughout this work. Simple modelling of non-linear strain fields confirms this conclusion.
	
	To best clarify the interplay of our phase-optimization approach and the physical quantities such as phase-jump and non-linear strain, in Fig.~S\ref{fig:s_strain} we include error bars on the extracted values of the phase jump. These error bars represent the statistical error, obtained in principle by adding in quadrature two relative errors: one from the spatial variation of the phase field (an error quantified by standard deviation of the phase field across both regions), and the other from optimizing $\mathbf{K}_\alpha$ (quantified by the fit-error in finding the minimum value of $\varepsilon_{\phi\alpha}(\mathbf{K}_\alpha)$), the latter being negligibly small throughout this work. These error bars include some of the uncertainty due to the spatial fluctuations of strain, but they do not at all contain the systematic error made by the method one chooses to remove the underlying smooth non-linear strain profile. As the example in the bottom row of Fig.~S\ref{fig:s_strain} demonstrates, when we extract the phase jump using three methods, namely, considering only the $R$ region, only the $L$ region, or both, we get three values of phase jump magnitude which agree within the statistical error bars. This is strong indication that non-linear strain is negligible in this example, i.e. we do not have a systematic error due to strain. The example in top row of Fig.~S\ref{fig:s_strain} produces three values of phase jump that do not overlap within the statistical error bars. Hence, we know that the three methods ($R$ only, $L$ only, or both) give three distinguishable reference lattice which compensate differently for the non-linear strain. As argued in previous paragraph, the method using both $L$ and $R$ will minimize the bias in phase jump, and gives a value in between the ones obtained from only $L$ or only $R$. Hence, for the few fields-of-view which are near a strained area, e.g. top row of Fig.~S\ref{fig:s_strain}, we report the result of our usual method that uses both $L$ and $R$.
	
	Finally, it is worth noting that using a reference lattice which is strictly integer-valued in pixels of the Fourier space is simply a systematic error which is more problematic, since it affects all phase jumps (not only near heavily strained areas), and may drastically bias the phase jump (able to move it from $\sim$$\pi/2$ to $\pi$) simply because a shift of $\mathbf{K}_\alpha$ by half a pixel corresponds to a spurious accumulation of phase of order $\pi$ across the field of view that may easily bias the extracted phase jump at the defect (see the schematic in main text Fig.~2a).
	
	\section{Dip-like 1D defect}
\begin{figure}
		\centering
		\includegraphics[width=0.95\textwidth]{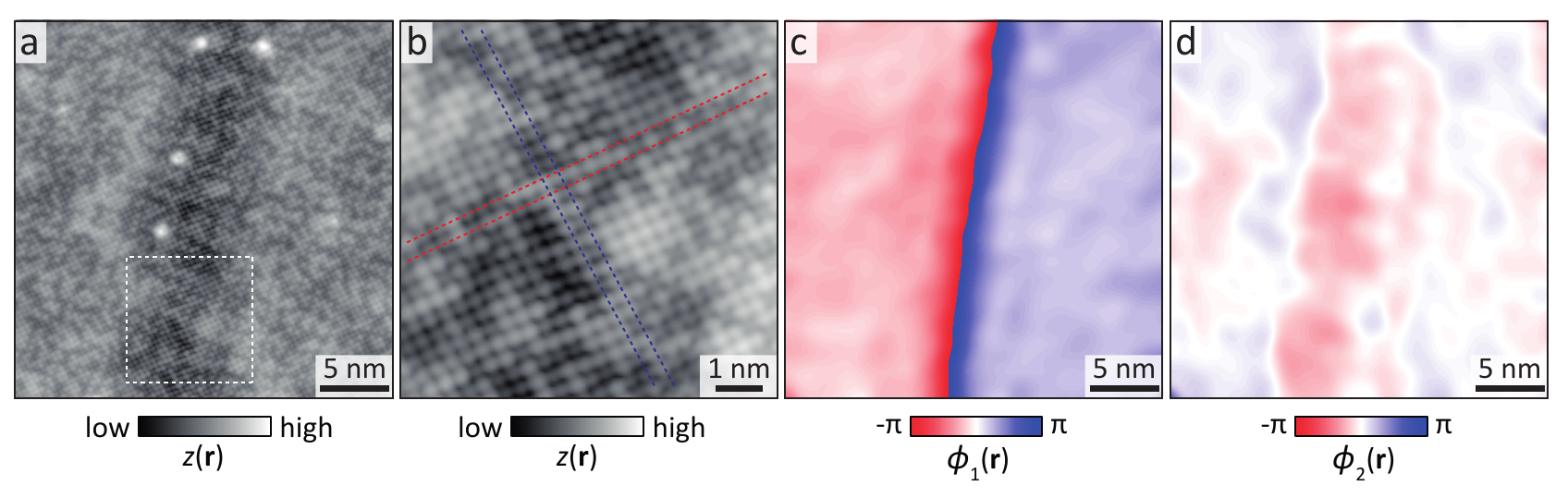}\\		
		\caption{\textbf{Dip-like defect: phase.} \textbf{a} Constant current image of a segment of the dip-like 1D defect, the dip is several nm wide and runs near-vertical. Setup: $V$ = 5~mV, $I$ = 50~pA. \textbf{b} Enlargement of the dashed box in panel \textbf{a} showing clear atomic contrast at the defect. The red and blue dashed lines are guides to the eye highlighting the shift of the atoms in the vertical direction upon crossing the defect. \textbf{c}, \textbf{d} Phase images for the near vertical and near horizontal Bragg peaks, respectively. The total magnitude of the phase shift is $\Delta \phi$ = 0.46$\pi\ \pm$ 0.1$\pi$.}
		\label{fig:s_dip_phase}
	\end{figure}

	In addition to the ridge-like 1D defect that is the focus of the main text, we found another 1D defect that appears as a depression in topography (i.e. it is dip-like). Figure S\ref{fig:s_dip_phase}a shows a segment of this defect. Unlike the ridge-like defect, all atoms are clearly visible in the dip-like defect, as evidenced from  the enlargement in Fig.~S\ref{fig:s_dip_phase}b. As the guides to the eye highlight, there is a clear shift in the atomic lattice in the vertical direction upon crossing the defect. Using the same phase analysis as for the ridge-like defect, we extract the phase shift in the direction of the two Bragg peaks, see Figs.~S\ref{fig:s_dip_phase}c,d. The magnitude and angle with respect to the defect for several locations along the hundreds of nm long defect are plotted in main text Fig.~2f: the magnitude is similar to that of the ridge-like defect ($\sim$$\pi /2$), but the shift direction is more along the defect instead of perpendicular to it as is the case for the ridge-like defect.
	
\begin{figure}
		\centering
		\includegraphics[width=\textwidth]{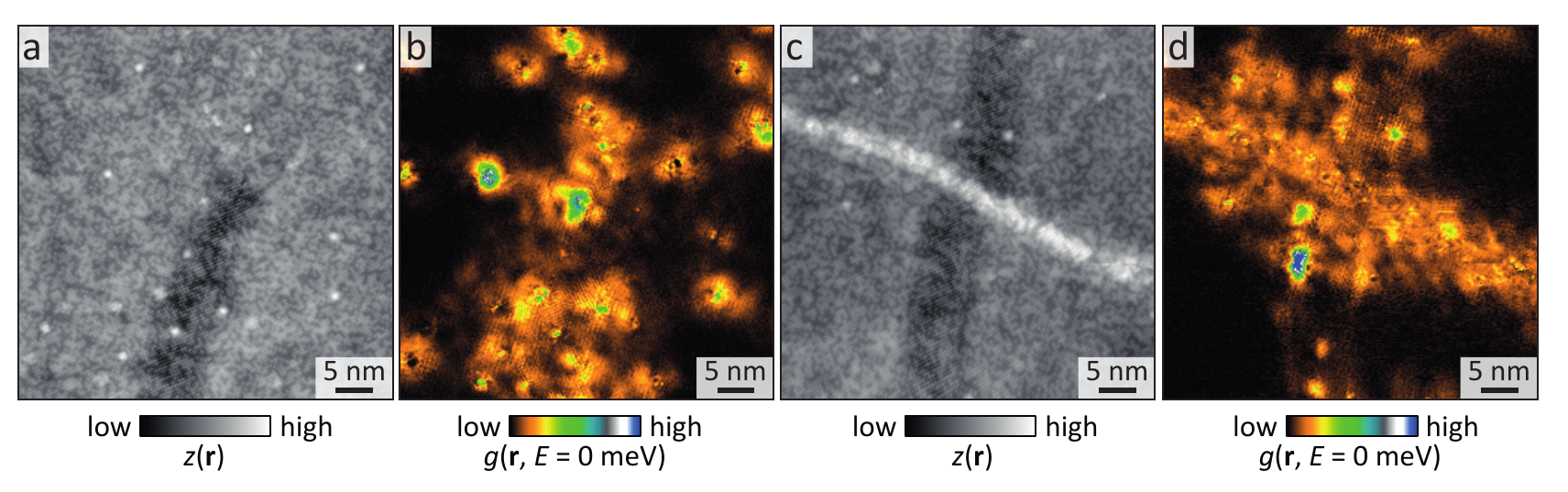}
		\caption{\textbf{Dip-like defect: conductance} \textbf{a} Constant current and \textbf{b} zero bias conductance at the endpoint of the dip-like 1D defect. \textbf{c}, \textbf{d} Same as panels \textbf{a}, \textbf{b}, but for a location where the dip-like defect crosses a ridge-like defect. Despite a similar phase shift of the lattice, the dip-like 1D defect shows hardly any sub-gap conductance, if at all. Setup conditions for both measurements: $V$ = 5~mV, $I$ = 100~pA}
		\label{fig:s_dip_didv}
	\end{figure}

	Since the dip-like defect also generates a phase shift of the lattice, which is of the same order of magnitude as that for the ridge-like defect, the question is whether there is any signature in differential conductance that would suggest the presence of topological non-trivial states. Figure \ref{fig:s_dip_didv} shows two examples that illustrate the absence of such signatures. The zero bias conductance at the endpoint of the dip is only showing intensity at excess iron impurities, with little to no contribution from the 1D defect. The marked contrast between the zero bias conductance of the dip-like defect and the ridge-like defect can be seen in the crossing of the two in Figs.~\ref{fig:s_dip_didv}c,d: whereas a strong sub-gap intensity is apparent on the latter, the former is barely, if at all, visible.
	
	\section{Surface debris}
\begin{figure}
		\centering
		\includegraphics[width=\textwidth]{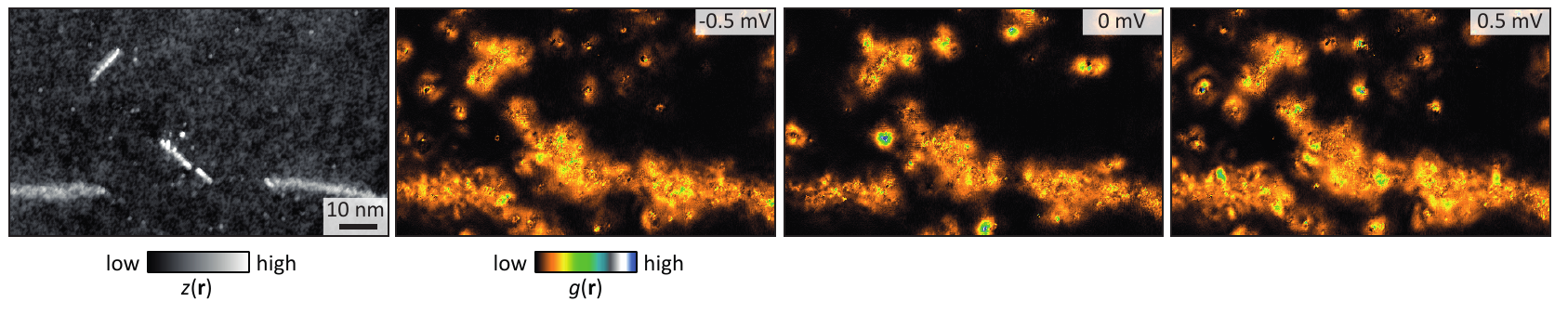}
		\caption{\textbf{Gaps and debris.} \textbf{a} Constant current image and \textbf{b}-\textbf{d} simultaneously recorded differential conductance at three different voltages (-0.5~mV, 0~mV, 0.5~mV). Both at zero bias and at other sub-gap voltages the 1D defect and surface debris are nearly indistinguishable in differential conductance. For the surface debris, no phase shift of the lattice is present. Main text Figs.~3a and b are the lower parts of the topography and zero bias conductance shown here, respectively. Setup: $V$ = 5~mV, $I$ = 100~pA.}
		\label{fig:s_junk_large}
	\end{figure}

	Main text Figs.~3a,b show a region where a long 1D defect is broken up: two internal endpoints are seen some tens of nanometres apart. Interestingly, in between the endpoints a strip of debris is located. As can be seen in the full field of view data in Fig.~S\ref{fig:s_junk_large}, a second, smaller strip of debris is located a little farther from the gap in the 1D defect. If the two strips of debris are laid end-to-end, they are not far from bridging the gap between the two internal endpoints, almost suggesting they once did, but were somehow ejected. The striking similarity in both shape (height, width and one-dimensionality) and sub-gap filling strongly suggests the two objects are actually the same: in one case residing on top of the surface (the debris, causing no lattice shift) and in the other case below the top layer (the ridge-like 1D defect, causing a lattice shift in the warped top layer).
	
\begin{figure}
		\centering
		\includegraphics[width=\textwidth]{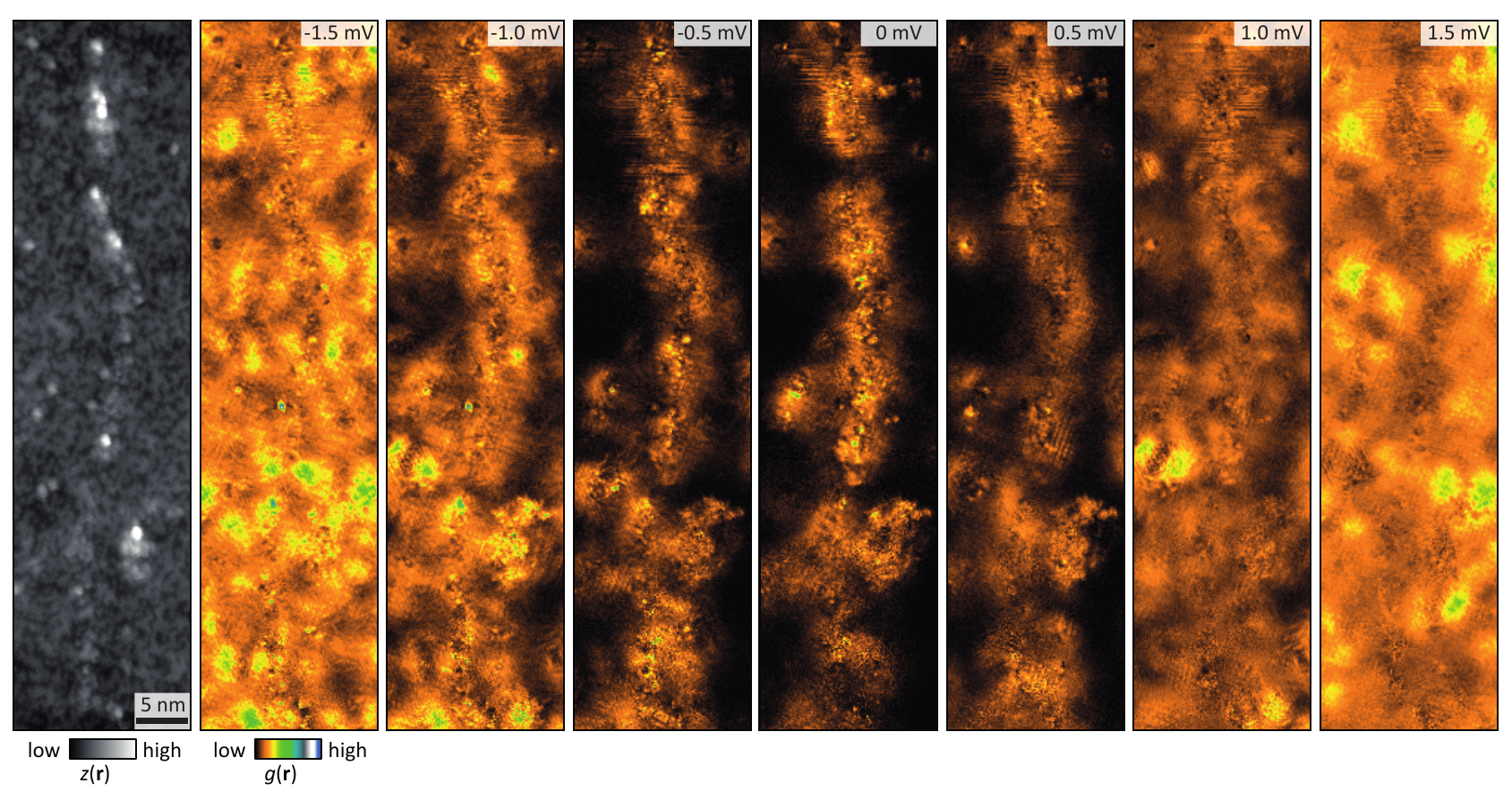}
		\caption{\textbf{Surface debris: gap filling}. Constant current image (left) and simultaneously recorded differential conductance (right) ranging from the negative to positive gap edge of a string of surface debris. Setup: $V$ = 5~mV, $I$ = 80~pA. The gap filling is near identical to that of the ridge-like 1D defects, yet is clearly of topologically trivial origin as the debris is disconnected and is not accompanied by a phase shift of the lattice.}
		\label{fig:s_junk}
	\end{figure}

	To highlight that the sub-gap filling along the ridge-like 1D defects is not unique to these objects, Fig.~S\ref{fig:s_junk} shows another example of surface debris where in absence of a phase shift of the lattice a near identical gap filling is observed (see e.g. Fig.~S\ref{fig:s_endpoint} and S\ref{fig:s_shortone} for comparison).
	
	\section{Conductance gaps along the 1D defect}
\begin{figure}
		\centering
		\includegraphics[width=\textwidth]{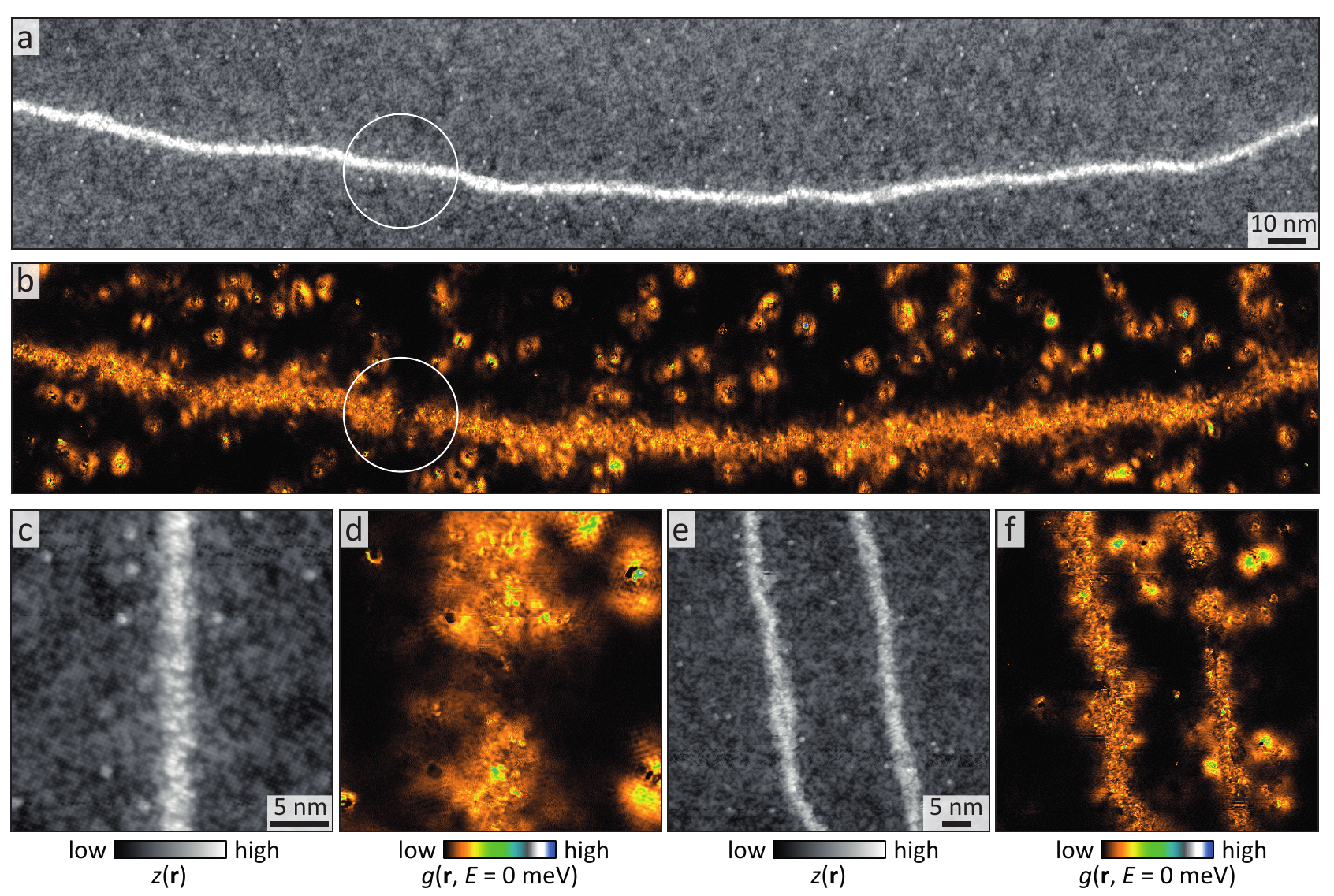}
		\caption{\textbf{Gaps in zero bias conductance.} \textbf{a} Constant current and \textbf{b} zero bias conductance of a large stretch of ridge-like 1D defect. Whereas the zero bias conductance is enhanced along the majority of the defect, a small region (indicated by the circle) is not. \textbf{c}, \textbf{d} enlargement of the gapped region in panels \textbf{a}, \textbf{b} respectively. \textbf{e}, \textbf{f} Another example of gapping on the 1D defect, in this case on one of two 1D defects that run parallel to each other (right-top). For all panels the setup conditions are $V$ = 5~mV, $I$ = 100~pA.}
		\label{fig:s_gaps}
	\end{figure}

	Main text Figs.~3d,e shows a location on one of the ridge-like 1D defects where a break in the sub-gap differential conductance is observed. Although such locations are rare, we found several other cases. Figs.~S\ref{fig:s_gaps}a,b show one example where the hundreds-of-nanometres-long 1D defect is continuous, yet around a certain location a gap is recovered in the differential conductance. The enlargement of this area in Figs.~S\ref{fig:s_gaps}c,d illustrates that the 1D defect itself does not look different in topography, whereas the zero-bias conductance is clearly vanishing due to a recovery of the gap in the density of states. Another example is shown in Figs.~S\ref{fig:s_gaps}e,f where two 1D defects are running parallel to each other and looking the same, while one of them has two short segments in which the zero-bias conductance vanishes. In all cases, the phase shift of the lattice across these segments does not differ from the rest of the 1D defect. This behaviour is not compatible with topologically protected modes.
	
	\section{Manipulation}
\begin{figure}
		\centering
		\includegraphics[width=\textwidth]{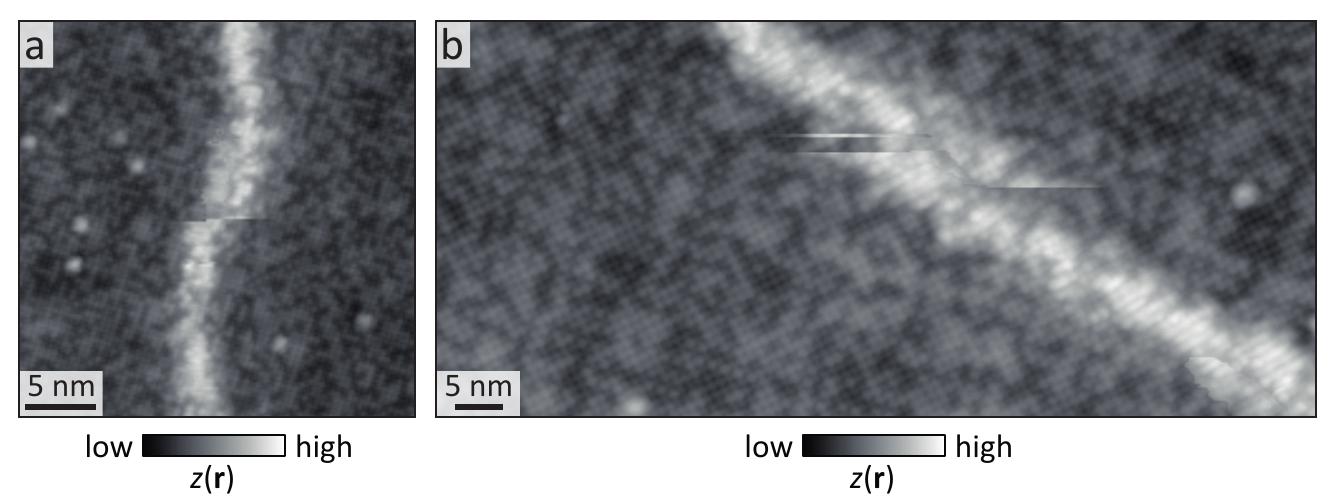}
		\caption{\textbf{Temporarily moving 1D defect}. \textbf{a}, \textbf{b} Constant current images on two different 1D defects. During imaging, both 1D defects move sideways and suddenly jump back. The atomic lattice is continuous across the line where the 1D defect jumps, showing that only the 1D defect is moved while the tip and surface remain the same. The y-axis is the slow scanning direction. Setup for both images: $V$ = 5~mV, $I$ = 100~pA.}
		\label{fig:s_temp_jump}
	\end{figure}

\begin{figure}
		\centering
		\includegraphics[width=\textwidth]{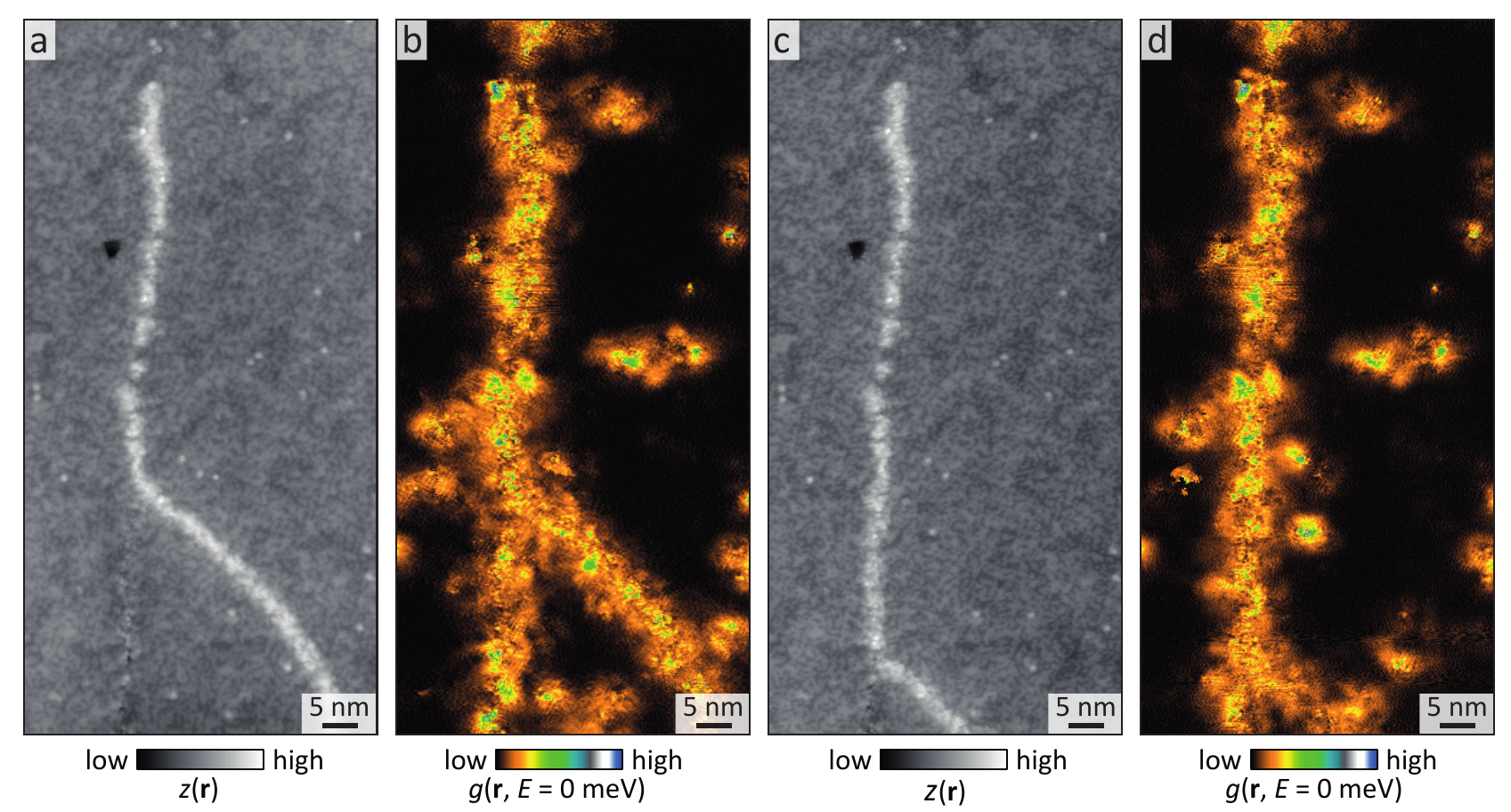}
		\caption{\textbf{Permanently moving 1D defect}. \textbf{a} Constant current image and \textbf{b} simultaneously recorded zero bias conductance before manipulation. \textbf{c}, \textbf{d} The same as panels \textbf{a}, \textbf{b}, respectively, after manipulation. Setup: $V$ = 5~mV, $I$ = 80~pA.}
		\label{fig:s_manipulation}
	\end{figure}

	As was shown in main text Fig.~4, we managed to move one of the 1D defects. Although we have observed instabilities on multiple 1D defects (at moderate junction resistances, $\sim$50~M$\Omega$), in all cases the defect snapped back to its original location. Two examples of such a temporary displacement are shown in Fig.~S\ref{fig:s_temp_jump}. These measurement clearly show that only the 1D defect is manipulated, without affecting the tip or surface: the lattice is continuous across the line where the 1D defect snaps back. We note that even for high junction resistances often spectra taken on 1D defects (as well as on surface debris) show enhanced noise with respect to those taken on defect free areas, as exemplified by the spectra in main text Fig.~3c. This is directly related to the local instability of the defects or sections thereof to the presence of the tip. The case of Fig.~4 was the only instance where the manipulation of a large section of 1D defect was permanent: it did not snap back again and remained in its new location for the remainder of our study. Figure~S\ref{fig:s_manipulation} shows a larger field of view measurement before and after manipulation of the same 1D defect as that shown in Fig.~4. To achieve this permanent manipulation, we scanned the tip at a high speed (120~nm/s versus our usual speed of $\sim$8~nm/s) and at a relatively low junction resistance (2.5~M$\Omega$). Of course at such settings there is unfortunately always a risk of altering the tip and/or surface as well. In this case, one excess Fe atom was inadvertently picked up by the tip (and later dropped elsewhere again): the atomic structure, including the large hole and all other excess Fe atoms were not affected by the manipulation.
	
	As further discussed in the next section, the differential conductance at the endpoint of the 1D defect (top of Fig.~S\ref{fig:s_manipulation}) is unremarkable. More importantly, in addition to enhanced sub-gap density of states on the ridge-like 1D defect, the differential conductance is also enhanced on another section that does not have a ridge-like 1D defect, forming an inverted Y in Fig.~S\ref{fig:s_manipulation}b. This section does not show a phase shift of the lattice, but actually appears damaged: atoms are missing and mixed in with point defects (see also main text Fig.~4a). It is most likely this damage that manages to pin the 1D defect in place once the electric field of the tip has dislocated it from its original position. In a series of measurements, we were indeed able to shift the 1D defect step-by-step, each time snapping a new section to the line of damage. This shows that the 1D defect is a flexible, movable object, not unlike the surface debris. 
	
	\section{End points}
\begin{figure}
		\centering
		\includegraphics[width=\textwidth]{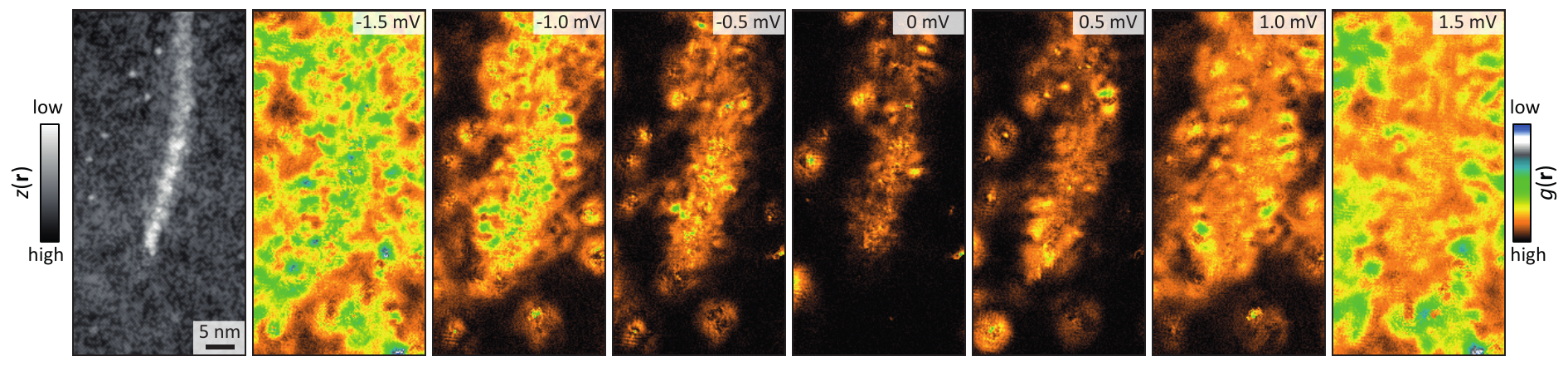}
		\caption{Constant current image (left) and simultaneously recorded differential conductance (right) ranging from the negative to positive gap edge of an endpoint of a 1D defect that is hundreds of nm long. Setup: $V$ = 5~mV, $I$ = 100~pA.}
		\label{fig:s_endpoint}
	\end{figure}

\begin{figure}
		\centering
		\includegraphics[width=0.8\textwidth]{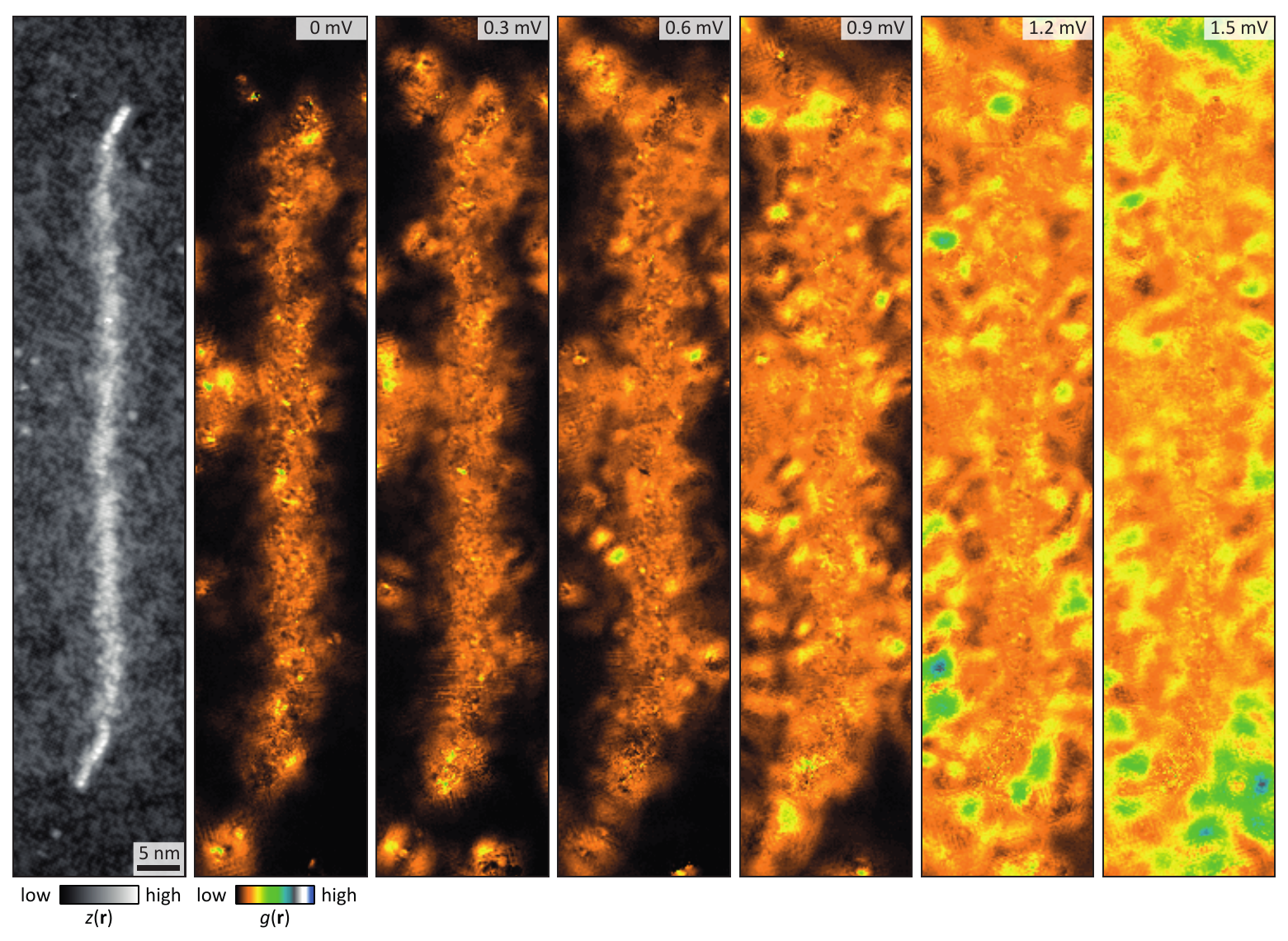}
		
		\caption{Constant current image (left) and simultaneously recorded differential conductance (right) ranging from zero bias to the positive gap edge of a relatively short 1D defect ($\sim$80~nm). Setup: $V$ = 5~mV, $I$ = 100~pA.}
		\label{fig:s_shortone}
	\end{figure}

	As discussed in the main text, even though based on our analysis the 1D defects do not host topologically non-trivial dispersing modes, their end-points may still host non-trivial modes, similar to helical spin chains on a superconductor. As main text Figs.~3a,b show, however, there does not appear to be a remarkable change in density of states at the ends of the defects. To illustrate this in more detail, Figs.~S\ref{fig:s_endpoint} and S\ref{fig:s_shortone} show the evolution of the differential conductance on the endpoints of two other 1D defects. Neither the end of a long defect (Fig.~S\ref{fig:s_endpoint}, hundreds of nm) or both ends of a relatively short one (Fig.~S\ref{fig:s_shortone}, $\sim$80~nm) are different from the midsection of the defect. Note also the striking resemblance of either data-set to that of the string of debris in Fig.~S\ref{fig:s_junk}.
	
\end{document}